\documentclass[11pt,fleqn]{article}

\usepackage{graphicx}
\usepackage{epsf}
\usepackage{fullpage}
\usepackage{amssymb}
\usepackage{latexsym}
\usepackage{amsmath}

\DeclareMathAlphabet{\bit}{OT1}{cmr}{bx}{sl}

\begin{document}

%\pagestyle{headings}
%\mainmatter
\newtheorem{theo}{Theorem}
\newtheorem{lemm}[theo]{Lemma}
\newtheorem{coro}[theo]{Corollary}
\newtheorem{obse}[theo]{Observation}
\newtheorem{prop}[theo]{Proposition}
\newtheorem{defi}[theo]{Definition}
\newtheorem{rema}[theo]{Remark}
\newtheorem{exam}[theo]{Example}
\newtheorem{conj}[theo]{Conjecture}
\newcommand{\demand}{\mbox{demand}}
\newcommand{\supply}{\mbox{supply}}
\newcommand{\proof}{\noindent\bf Proof: \rm}
\newcommand{\sketch}{\noindent\bf Sketch: \rm}
\newcommand{\qed}{$\Box$}
\newcommand{\eps}{\varepsilon}
\newcommand{\Rd}{{\mathbb{R}^{d}}}
\newcommand{\Rp}{{\mathbb{R}^{2}}}
\newcommand{\Rr}{{\mathbb{R}^{3}}}

\newcommand{\old}[1]{{}}

\title{The Geometric Maximum Traveling Salesman Problem\thanks{
Preliminary versions of parts of this paper appear in the Proceedings
of IPCO'98~\cite{BaJoWoWo98} and SODA'99~\cite{F99}.}}

%\titlerunning{The Maximum TSP}

\author{
Alexander Barvinok\thanks{Supported by an Alfred P. Sloan Research Fellowship
and NSF grant DMS 9501129.}\\
Dept.\ of Mathematics\\
University of Michigan\\
Ann Arbor, MI 48109-1009, USA\\
{\tt barvinok@math.lsa.umich.edu}\\
\and
S\'andor P.\ Fekete\thanks{
Partly supported by the Hermann-Minkowski-Minerva
Center for Geometry at Tel Aviv University, while visiting
the Center in March 1998; other parts 
supported by the Deutsche Forschungsgemeinschaft, FE 407/3-1,
when visiting Rice University in June 1998.}\\
Dept.\ of Mathematical Optimization\\
Braunschweig University of Technology\\
38106 Braunschweig, GERMANY \\
{\tt s.fekete@tu-bs.de}\\
\and David S. Johnson\\
AT\&T Research\\
AT\&T Labs\\
Florham Park, NJ 07932-0971, USA\\
{\tt dsj@research.att.com}\\
\and Arie Tamir\\
School of Mathematical Sciences\\
Tel Aviv University\\
Tel Aviv, ISRAEL\\
{\tt atamir@math.tau.ac.il}\\
\and Gerhard J. Woeginger\thanks{
Supported by the START program Y43-MAT of the Austrian Ministry of Science.}\\
Dept.\ of Mathematics\\
University of Twente,
P.O. Box 217\\ 7500 AE Enschede,
THE NETHERLANDS\\ 
{\tt g.j.woeginger@math.utwente.nl }
\and Russ Woodroofe\thanks{
Supported by the NSF through the REU Program while at the Dept.\
of Mathematics, University of Michigan.}\\
Dept.\ of Mathematics\\
Cornell University\\
Ithaca, NY 14853-4201, USA\\
{\tt paranoia@math.cornell.edu }
}
%\authorrunning{Barvinok et al.}
\date{}
\maketitle

\begin{abstract}
We consider the traveling salesman problem when the cities are points 
in $\Rd$ for some fixed $d$ and distances are computed according to 
geometric distances, determined by some norm.  
We show that for any polyhedral norm, the problem of finding a tour of 
{\em maximum\/} length can be solved in polynomial time.
If arithmetic operations are assumed to take unit time, our algorithms run 
in time $O(n^{f-2}\log n)$, where $f$ is the number of facets of the polyhedron 
determining the polyhedral norm.  
Thus for example we have $O(n^2\log n)$ algorithms for the cases of points in 
the plane under the Rectilinear and Sup norms.
This is in contrast to the fact that finding a {\em minimum\/} length tour 
in each case is NP-hard. Our approach can be extended to the more general
case of {\em quasi-norms} with not necessarily symmetric unit ball,
where we get a complexity of $O(n^{2f-2}\log n)$.

For the special case of two-dimensional metrics with $f=4$
(which includes the Rectilinear and Sup norms),
we present a simple algorithm with $O(n)$ running time. 
The algorithm does not use any indirect addressing, so
its running time remains valid even in comparison based models
in which sorting requires $\Omega(n \log n)$ time. The basic 
mechanism of the algorithm provides some intuition on why
polyhedral norms allow fast algorithms.

Complementing the results on simplicity for polyhedral norms, 
we prove that for the case of Euclidean distances in $\Rd$
for $d\geq 3$, the Maximum TSP is NP-hard. This sheds new light
on the well-studied difficulties
of Euclidean distances.
\end{abstract}

\section{Introduction}
\label{sec:intr}

In the {\em Traveling Salesman Problem\/} (TSP), the input consists of a 
set $C$ of {\em cities\/} together with the 
distances $d(c,c')$ between every pair of distinct cities $c,c' \in C$.
The goal is to find an ordering or {\em tour\/} of the cities that minimizes
(Minimum TSP) or maximizes (Maximum TSP) the total tour length.
Here the length of a tour $c_{\pi(1)}, c_{\pi(2)}, \ldots ,c_{\pi(n)}$ is
\[
\sum_{i=1}^{n-1} d(c_{\pi(i)}, c_{\pi(i+1)}) + d(c_{\pi(n)},c_{\pi(1)}).
\]

Like the Minimum TSP, the Maximum TSP is NP-complete on graphs, even if
the triangle inequality holds.
After 25 years,
the best known performance guarantee for the metric Minimum
TSP is still Christofides's 3/2 approximation algorithm;
the results of
Arora, Lund, Motwani, Sudan, and Szegedy~\cite{pcp}
show that the problem cannot be approximated arbitrarily well,
combined with results by Papadimitriou and Yannakakis~\cite{py},
it follows that this holds even for the special class
of instances where all distances
are 1 or 2. There has been more
development on approximation algorithms for the metric Maximum TSP:
Recently, Hassin and Rubinstein~\cite{7/8} have given a 7/8
approximation algorithm.

Of particular interest are {\em geometric\/} instances of the TSP,
in which cities correspond to points in $\Rd$ for some $d \geq 1$,
and distances are computed according to some geometric norm.
Perhaps the most popular norms are the Rectilinear, Euclidean, and
Sup norms.
These are examples of what is known as an ``$L_p$ norm''
for $p = 1$, $2$, and $\infty$.
In general, the distance between two points
${\bit x} = (x_1,x_2,\ldots,x_d)$ and
${\bit y} = (y_1,y_2,\ldots,y_d)$ under the $L_p$ norm, $p \geq 1$, is
\[
d({\bit x} , {\bit y} ) = 
\left( \sum_{i=1}^d \left| x_i - y_i \right|^p \right) ^{1/p}
\]
with the natural asymptotic interpretation that distance under the $L_\infty$ norm
is 
\[
d({\bit x} , {\bit y} ) = 
\max{}_{i=1}^d \left|x_i - y_i\right|.
\]
This paper considers a second class of norms which also includes
the Rectilinear and Sup norms, but can only approximate the Euclidean
and other $L_p$ norms.
This is the class of {\em polyhedral norms\/}.
Each polyhedral norm is determined by a {\em unit ball\/} which is a
centrally-symmetric polyhedron ${\bit P}$ with the origin at its center.
To determine $d({\bit x} , {\bit y} )$ under such a norm,
first translate the space so that one of the points, say ${\bit x}$,
is at the origin.
Then determine the unique factor $\alpha$ by which one must rescale
${\bit P}$ (expanding if $\alpha > 1$, shrinking if $\alpha < 1$) so that
the other point (${\bit y}$) is on the boundary of the polyhedron.
We then have $d({\bit x} , {\bit y} ) = \alpha$.

Alternatively, and more usefully for our purposes, we can view a polyhedral
norm as follows.
If ${\bit P}$ is a polyhedron as described above and has $f$ facets, then
$f$ is divisible by 2 and there is a set 
$H_{\bit P} = \{{\bit h}_1,\ldots ,{\bit h}_{f/2}\}$ of points in $\Rd$
such that ${\bit P}$ is the intersection of a collection of half-spaces 
determined by $H_{{\bit P}}$:
\[
{\bit P} = 
\left( \bigcap_{i=1}^{f/2} \{{\bit x}:~
{\bit x} \cdot {\bit h}_i \leq 1 \} \right ) 
\cap 
\left( \bigcap_{i=1}^{f/2} \{{\bit x}:~ 
{\bit x} \cdot {\bit h}_i \geq -1 \} \right).
\]
Then we have
\[
d({\bit x} , {\bit y} ) = \max \Big \{ \Big |({\bit x}-{\bit y}) 
\cdot {\bit h}_i \Big |:~ 1 \leq i \leq {f/2} \Big \}. 
\]
Note that for the Rectilinear norm in the plane we can take $H_{{\bit P}}$ 
= $\{(1,1),(-1,1)\}$ and for the Sup norm in the plane we can take 
$H_{{\bit P}}$ = $\{(1,0),(0,1)\}$.

%It is possible to consider asymmetric distance functions and charaterize them
%by their unit ball or a set of vectors. As we will see, these {\em quasi-norms}
%can be discussed quite similarly to regular symmetric norms.

\medskip
For the Minimum TSP on geometric instances, two key complexity 
questions have been answered. 
As follows from results of Itai, Papadimitriou, and Swarcfiter 
\cite{ItPaSw}, the Minimum TSP is NP-hard for any fixed dimension $d$ 
and any $L_p$ or polyhedral norm; see also the earlier results by
Garey, Graham, and Johnson~\cite{GGJ}, and Papadimitriou~\cite{Pap}.
On the other hand, results of Arora \cite{arora1} and 
Mitchell \cite{mitch} imply that in all these cases a polynomial-time 
approximation scheme (PTAS) exists, i.e., a sequence of polynomial-time 
algorithms $A_k$, $1 \leq k < \infty$, where $A_k$ is guaranteed to find 
a tour whose length is within a ratio of $1+(1/k)$ of optimal.

The situation for geometric versions of the Maximum TSP has been less
clear than for its minimum counterpart. Serdyukov~\cite{Serd2},
and, independently, Barvinok \cite{Barvinok}, 
have shown that once again polynomial-time
approximation schemes exist for all fixed dimensions $d$ and all
$L_p$ or polyhedral norms (and in a sense for {\em any\/} fixed norm;
see \cite{Barvinok}). 
Until now, however, the complexity of the optimization problems themselves
when $d$ is fixed has remained open: For no fixed dimension $d$ and $L_p$ 
or polyhedral norm was the problem of determining the maximum tour length 
known either to be NP-hard or to be polynomial-time solvable.
%(Unknown to the authors of this paper, Serdyukov~\cite{Serd2,Serd3} showed
%that for the case of a quasi-norm with a triangle as the unit ball,
%the Maximum TSP can be solved in polynomial time.)
In this paper, we resolve the question for all polyhedral norms, showing that, 
in contrast to the case for the Minimum TSP, the Maximum TSP is solvable
in polynomial time for any fixed dimension $d$ and any polyhedral norm:

\begin{theo} 
\label{th:main}
Let dimension $d$ be fixed, and let $\parallel\cdot\parallel$ be a fixed 
polyhedral norm in $\Rd$ whose unit ball is a centrally symmetric
polyhedron ${\bit P}$ determined by a set of $f$ facets.
Then for any set of $n$ points in $\Rd$, one can construct a traveling 
salesman tour of maximum length with respect to $\parallel\cdot\parallel$ 
in time $O(n^{f-2}\log n)$ in the real number RAM model, where
arithmetic operations take unit time and only addition,
subtraction, multiplication, and division are allowed.
\end{theo}

%With an adjusted running time of $O(n^{2f-2}\log n)$,
%this result remains valid for polyhedral quasi-norms.

A similar result also holds for polyhedral {\em quasi-norms}, i.e., asymmetric
distance functions that can be defined in terms of ``unit balls'' that are
non-symmetric polyhedra containing the origin.  These can again be characterized
by a set of vectors $H_P$, but now we need a vector for each facet of
the polyhedron and
$d(x,y) = \max \left\{(x-y)\cdot h~:~h \in H_P \right\}$.
Serdyukov \cite{Serd2,Serd3}
previously showed that for the case of a quasi-norm in $\Rp$
with a triangle as the unit ball, the Maximum TSP can be solved in polynomial
time.  With our techniques we can show the following analog of
Theorem \ref{th:main}.

\begin{theo}
\label{th:quasi}
Let dimension $d$ be fixed, and let $\parallel\cdot\parallel$ be a fixed
polyhedral quasi-norm in $\Rd$ whose unit ball is a polyhedron ${\bit P}$
determined by a set of $f$ facets.
Then for any set of $n$ points in $\Rd$, one can construct a traveling
salesman tour of maximum length with respect to $\parallel\cdot\parallel$
in time $O(n^{2f-2}\log n)$ in the real number RAM model.
\end{theo}

As an immediate consequence of Theorem \ref{th:main}, we get relatively 
efficient algorithms for the Maximum TSP in the plane under Rectilinear 
and Sup norms, with a complexity of $O(n^2\log n)$.

%\begin{coro}
%\label{coro:rect}
%The Maximum TSP for points in $\Rp$ under the $L_1$ and $L_\infty$ 
%norms can be solved in $O(n^2\log n)$ time in the real number RAM model.
%\end{coro}

The restriction to the real number RAM model in Theorem \ref{th:main} 
is made primarily to simplify the statements 
of the conclusions. 
Suppose on the other hand that one assumes, as one typically must for 
complexity theory results, that the components of the vectors in 
$H_{\bit P}$ and the coordinates of the cities are all rationals.
Let $U$ denote the maximum absolute value of any of the corresponding
numerators and denominators.
Then the conclusions of the Theorem 
hold for the standard logarithmic cost RAM model
with running times multiplied by $n\log(U)$.
If the components/coordinates are all integers with maximum absolute 
value $U$, the running times need only be multiplied by $\log(nU)$.
For simplicity in the remainder of this paper, we shall stick to the
model in which numbers can be arbitrary reals and arithmetic operations
take unit time. The reader should have no trouble deriving the
above variants.

The above results make use of a polynomial
solution method
for the following TSP variant, which may be of independent interest: 
When visiting a given set
of $n$ cities, all connections have to be made
via a set of $k$ hubs, where $k$ is a constant.

The complexity of $O(n^2\log n)$ for the scenario of planar rectilinear
distances can be improved by
using special geometric properties of this case.
We can show the following optimal running time:

\begin{theo}
\label{th:linear}
The Maximum TSP for points in $\Rp$ under the $L_1$ 
norm can be solved in $O(n)$ time.
\end{theo}

By appropriate coordinate transformation, this result can 
be generalized to all planar polyhedral norms with $f=4$ facets,
which includes the Sup norm.
It holds even in a restricted model of computation,
where no indirect addressing may be used, and hence sorting
requires $\Omega(n \log n)$ time.
The main idea behind the algorithm
is to exploit the fact that rectilinear distances in the plane
have a high degree of degeneracy. As a consequence, we can 
show that the number of optimal tours is very large, 
$\Omega\left((\frac{n}{4}!)^4\right)$, and the set of 
optimal tours can be described very easily.
This contrasts sharply to the case of Euclidean distances,
where there may be a single optimal tour.

Indeed the complexity of the Maximum TSP in $\Rp$ under the Euclidean
metric remains an open question, although we have resolved the question
for higher dimensions with the following result.
\old{Despite of the variety of results on the Maximum TSP for
polyhedral distances, the case of Euclidean distances
in fixed-dimensional space
has been a main open problem. We resolve this problem
by proving that
the Maximum TSP with Euclidean distances
is NP-hard for any fixed $d\geq 3$, compounding the above
difficulties of Euclidean metrics by
a clear distinction in complexity between geometric instances 
with Euclidean and with rectilinear distances. 
}

\begin{theo}
\label{th:nphard}
Maximum TSP under Euclidean distances in $\Rd$ is an NP-hard
problem if $d\geq 3$.
\end{theo}

One of the consequences
is NP-hardness of the Maximum TSP for polyhedral norms with
an unbounded number of facets on the corresponding unit ball.
Another consequence concerns the so-called {\em Maximum Scatter TSP},
where the objective is to find a tour that maximizes the shortest edge.
The Maximum Scatter TSP was first considered by Arkin,
Chiang, Mitchell, Skiena, and Yang~\cite{ACMSY}, and the complexity
for geometric instances was stated as an open problem.
Our result implies NP-hardness for Euclidean
instances in 3-dimensional space.

An issue that is still unresolved for the Maximum TSP
as well as the Minimum TSP is the question of whether the
TSP under Euclidean distances is a member of the class NP,
allowing polynomial time verification of a good solution. 
Even if all city coordinates are rationals, we do not know how to compare
a tour length to a given rational target in less than
exponential time.
Such a comparison would appear to require us to evaluate a sum of $n$
square roots to some precision, and currently the best upper bound known
on the number of bits of precision needed to insure a correct answer
remains exponential in $n$. 

The rest of this paper is organized as follows. 
Section~\ref{sec:auxi} introduces a new special case of the TSP, the 
{\em Tunneling TSP\/}. We show how the Maximum TSP under a polyhedral 
norm can be reduced to the Tunneling TSP with the same number of
cities and $k=f/2$ tunnels (and with $k=f$ tunnels
for polyhedral quasi-norms).
Section~\ref{subsec:tunnel} shows how the solutions
for the Tunneling TSP with a fixed number $k\geq 2$ of tunnels
can be characterized, setting up 
an algorithm with a running time of $O(n^{2k-2}\log n)$
described in detail in Section~\ref{subsec:speedup}.
Section~\ref{sec:linear} describes the linear-time algorithm
for rectilinear distances in the plane. Section~\ref{sec:npc}
gives the NP-hardness proof of the Maximum Traveling Salesman
Problem for Euclidean distances in $\Rr$, and a number of higher-dimensional
extensions.
Section~\ref{sec:conc} concludes with a brief discussion and open problems.

\section{The Tunneling TSP}
\label{sec:auxi}
\nopagebreak
The {\em Tunneling TSP\/} is a special case of the Maximum TSP in which 
distances are determined by what we shall call a {\em tunnel system\/} 
distance function.
In such a distance function we are given a set 
$T = \{t_1,t_2, \ldots ,t_k\}$ of
$k\geq 2$ auxiliary objects that we shall call {\em tunnels\/}.
Each tunnel is viewed as a bidirectional passage having a front and a 
back end.
For each pair $c,t$ of a city and a tunnel we are given real-valued
{\em access distances\/} $F(c,t)$ and $B(c,t)$ from the city to the front 
and back ends of the tunnel respectively.
Each potential tour edge $\{c,c'\}$ must pass through some tunnel $t$,
either by entering the front end and leaving the back
(for a distance of $F(c,t)+B(c',t)$), or by entering
the back end and leaving the front (for a distance of $B(c,t)+F(c',t)$).
Since we are looking for a tour of maximum length, we can thus
define the distance between cities $c$ and $c'$ to be
\[
d(c,c') = \max \Big \{ F(c,t_i)+B(c', t_i),\;B(c,t_i)+F(c',t_i): 
1 \leq i \leq k \Big \}
\]
Note that this distance function, like our geometric norms, is symmetric.
(As we will see below, we can make adjustments
for asymmetric distance functions.)

It is easy to see that Maximum TSP remains NP-hard when distances are 
determined by arbitrary tunnel system distance functions.
However, for the case where $k = |T|$ is fixed and not 
part of the input, we will show in the next section that Maximum TSP
can be solved in $O(n^{2k-2}\log n)$ time.
We are interested in this special case because of the following lemma.
\begin{lemm} \label{le:tunnel}
If $\parallel\cdot\parallel$ is a polyhedral norm determined by a set
$H_{\bit P}$ of $k=f/2$ vectors in $\Rd$, then for any set $C$ of points in 
$\Rd$ one can in time $O(dk|C|)$ construct a tunnel system distance 
function with $k$ tunnels that yields
$d({\bit c},{\bit c'}) = \parallel {\bit c}-{\bit c'} \parallel$
for all ${\bit c},{\bit c'} \in C$.
\end{lemm}
\proof
The polyhedral distance between two cities 
$\parallel {\bit c},{\bit c'} \parallel \in \Rd$ is 
\begin{eqnarray*}
\parallel {\bit c}-{\bit c'} \parallel 
&=& 
\max \Big \{ \Big |({\bit c}-{\bit c'}) \cdot 
{\bit h}_i \Big |:~ 1 \leq i \leq k \Big \} \\
&=& 
\max \Big \{ ({\bit c}-{\bit c'}) \cdot 
{\bit h}_i,\; ({\bit c'}-{\bit c}) \cdot 
{\bit h}_i:~  1 \leq i \leq k \Big \}
\end{eqnarray*}
Thus we can view the distance function determined by 
$\parallel\cdot\parallel$ as a tunnel system distance function with set 
of tunnels $T = H_{\bit P}$ and 
$F({\bit c},{\bit h}) = {\bit c} \cdot {\bit h}$,
$B({\bit c},{\bit h}) = -{\bit c} \cdot {\bit h}$ for all cities
${\bit c}$ and tunnels ${\bit h}$.
\qed

\medskip
It is straightforward
to extend this characterization to the case of any polyhedral quasi-norm that
is characterized by a unit ball with a total of $f$ facets.
The only real change is in the complexity of the characterization
of the tunnel system in Lemma~\ref{le:tunnel}.
If we use the definition 
\[
\tilde{d}(c,c') = \max \Big \{ F(c,t_i)+B(c', t_i):
1 \leq i \leq k \Big \}
\]
for possibly asymmetric tunnel distances,
then by similar reasoning we get the following:

\begin{lemm} \label{le:quasi.tunnel}
If $\parallel\cdot\parallel$ is a polyhedral quasi-norm determined by a set
$H_{\bit P}$ of $f$ vectors in $\Rd$, then for any set $C$ of points in
$\Rd$ one can in time $O(df|C|)$ construct a tunnel system distance
function with $f$ tunnels that yields
$\tilde{d}({\bit c},{\bit c'}) = \parallel {\bit c}-{\bit c'} \parallel$
for all ${\bit c},{\bit c'} \in C$.
\end{lemm}

%\proof
%The polyhedral quasi-distance between two cities
%${\bit c},{\bit c'} \in \Rd$ is
%%
%\begin{eqnarray*}
%\parallel {\bit c}-{\bit c'}\parallel  
%&=&
%\max \Big \{ ({\bit c}-{\bit c'}) \cdot
%{\bit h}_i :~ 1 \leq i \leq f \Big \}. 
%\hspace{.1in}\Box
%\end{eqnarray*}

\newpage
\section{An Algorithm for Bounded Tunnel Systems}
\label{sec:main}
In this section we describe an $O(n^{2k-2} \log n)$ algorithm to 
solve the Tunneling TSP when the number of
tunnels is fixed at $k$,
assuming the real number RAM model.
By Lemmas \ref{le:tunnel} and \ref{le:quasi.tunnel} this implies
our results for polyhedral norms and quasi-norms
(Theorems \ref{th:main} and \ref{th:quasi}.)
The approach described below also yields a polynomial algorithm for 
a Minimum TSP variant, where a given set
of $n$ cities has to be traveled, and all connections have to be made
via a set of a fixed number $k$ of hubs.

We start by characterizing solutions for bounded tunnel systems
in Section~\ref{subsec:tunnel}.
This characterization is the basis
for our algorithm, which is described  in Section~\ref{subsec:speedup}.
In Section~\ref{subsec:further} we present an
additional  idea that may possibly improve the above
complexity to $O(n^{2k-2})$.
%Finally, we describe how to extend our results to quasi-norms.

\subsection{Characterizing Solutions for Bounded Tunnel Systems}
\label{subsec:tunnel}
We start by discussing the solutions for bounded tunnel systems.

Suppose we are given an instance of the Tunneling TSP with sets
$C = \{c_1,\ldots,c_n\}$ and $T=\{t_1,\ldots,t_k\}$ of cities
and tunnels, and access distances $F(c,t)$, $B(c,t)$ for all $c \in C$
and $t \in T$.
We begin by transforming the problem to one about subset construction.

Let $G=(C \cup T,E)$ be an edge-weighted, bipartite multigraph
with four edges between each city $c$ and tunnel $t$,
denoted by $e_i[c,t,X]$, $i \in \{1,2\}$ and $X \in \{B,F\}$.
The weights of these edges are $w(e_i[c,t,F]) = F(c,t)$
and $w(e_i[c,t,B]) = B(c,t)$, $i \in \{1,2\}$.
For notational convenience, let us partition the edges in $E$ into
sets $E[t,F] = \{e_i[c,t,F]: c \in C, i \in \{1,2\}\}$ and
$E[t,B] = \{e_i[c,t,B]: c \in C, i \in \{1,2\}\}$,
$t \in T$.
Each tour for the TSP instance then corresponds to a subset $E'$ of
$E$ that has $\sum_{e \in E'} w(e)$ equal to the tour length and satisfies
\begin{quote}
\begin{itemize} \itemsep=0.2ex
\item[(T1)]
Every city is incident to exactly two edges in $E'$.
\item[(T2)]
For each tunnel $t \in T$, $|E' \cap E[t,F]| = |E' \cap E[t,B]|$.
\item[(T3)]
The set $E'$ is connected.
\end{itemize}
\end{quote}

\noindent
To construct the multiset $E'$, we simply represent each tour edge
$\{c,c'\}$ by a pair of edges from $E$ that connect in the appropriate
way to the tunnel that determines $d(c,c')$.  For example,
if $d(c,c') = F(c,t)+B(c',t)$, and $c$ appears immediately before $c'$
when the tour is traversed starting from $c_{\pi (1)}$, then the edge
$(c,c')$ can be represented by the two edges $e_2[c,t,F]$ and $e_1[c',t,B]$.
Note that there are enough (city,tunnel) edges of each type so that all
tour edges can be represented, even if a given city uses the same tunnel
endpoint for both its tour edges.
Also note that if
$d(c,c')$ can be realized in more than one way, the multiset $E'$ will
not be unique.
However, any multiset $E'$ constructed in this fashion will still have
$\sum_{e \in E'} w(e)$ equal to the tour length.

On the other hand, any set $E'$ satisfying (T1) -- (T3) corresponds to one
(or more) tours having length at least $\sum_{e \in E'} w(e)$:
Let $T' \subseteq T$ be the set of tunnels $t$ with $|E' \cap E[t,F]| > 0$.
Then $G' = (C \cup T', E')$ is a connected graph
all of whose vertex degrees are even by (T1) -- (T3).
By an easy result from graph theory, this means that $G'$ contains
an Euler tour that by (T1) passes through each city exactly once,
thus inducing a TSP tour for $C$.
Moreover, by (T2) one can construct such an Euler tour with the additional
property that if $e_i[c,t,x]$ and $e_j[c',t,y]$
are consecutive edges in this tour, then $x \neq y$, i.e, either
$x=F, y=B$ or $x=B, y=F$.
Thus we will have $w(e_i[c,t,x])+w(e_j[c',t,y]) \leq d(c,c')$, and hence the
length of the TSP tour will be at least $\sum_{e \in E'} w(e)$, as claimed.

In summary, our problem is reduced to finding a maximum weight set of edges
$E' \subseteq E$ satisfying (T1) -- (T3).

\vspace{.1in}
\subsection{An Efficient Algorithm}
\label{subsec:speedup}

\subsubsection{Identifiers for Subproblems}
\label{subsubsec:identifiers}

{From} the previous section we know that our problem
is reduced to finding a maximum weight set of edges
$E'\subseteq E$ satisfying (T1) -- (T3).
Given that $k$ is fixed, we will first show how to decompose the problem into $O(n^{2k-3})$ instances,
and then demonstrate how to solve each one of them in $O(n \log n)$ time.

Let $E'$ be a subset of edges satisfying (T1) -- (T3).
We associate five identifiers with $E'$.

The first is $T'\subseteq T$,
the subset of $p$, $p \le k$, tunnels that are spanned by $E'$.
(Without loss of generality suppose that $T'=\{t_1,...,t_p\}$.)
Globally, the total number of identifiers of the first type is
$O(2^k)$.

Due to conditions (T1), (T3) we know that there is a subset of
$2(p-1)$ edges $E''\subset E'$ such that
the subgraph $G(E'')$, induced by $E''$, is connected and
it spans $T'$ and exactly $p-1$ cities.
Moreover, each of these cities
is incident to exactly two edges in $E''$, which connects two
adjacent tunnels in $T'$.  We use the second, the
third and the fourth identifiers to characterize $G(E'')$.

The second identifier is the spanning tree topology
connecting the tunnels of $T'$ that is induced
by the set of edges $E''$. There are $p^{p-2}$ identifiers
of this type: the number of labeled spanning trees of $K_p$,
the complete graph on $p$ nodes \cite{Cayley}.

The third identifier is an assignment of distinct cities to the
spanning tree edges of the previous identifier.  The city assigned
to the tree edge joining tunnels $t$ and $t'$ is one that is connected
by one edge of $E''$ to $t$ and by another edge of $E''$ to $t'$.
The total number of identifiers of the third type is $O(n^{p-1})$.

The fourth identifier is the subset of $2(p-1)$ edges
linking these cities with corresponding  tunnel entrances.
In the underlying graph $G$ there are  four edges (two pairs of
identical edges) connecting a city with a tunnel. Therefore,
the total number of identifiers of the fourth type is
$O(16^{p-1})$.

The fifth and last identifier is the degree sequence
of the nodes in $T'$ induced by $E'$. Let $d=(2d_1,...,2d_p)$ denote
this sequence of even positive degrees. Note that $d_1+...+d_p=n$.
Therefore, there are clearly
$O(n^{p-1})$ identifiers of this third type. However,
we will modify the identifier and use one of the
degree entries, say $d_p$, as an unspecified parameter. Thus,
there are only $O(n^{p-2})$ choices
for, say, $d_1,d_2,...,d_{p-2}$.
($d_{p-1}$ will then depend linearly on  the parameter $d_p$.)
Altogether, we now
have $O(n^{2k-3})$ choices for values of identifiers.

In summary, to prove our claim that the Tunneling TSP can be
solved in $O(n^{2k-2} \log n)$ time
when $k$ is fixed, it will suffice to show that the following
problem can be solved in $O(n \log n)$ time.

Given a set of tunnels $T'=\{t_1,...,t_p\}$, a set of $p-1$ cities,
a set $E''$ of $2(p-1)$ edges connecting the cities to the tunnels
as above, and a set of positive integers $d_1,...,d_{p-2}$,
find a maximum weight set of edges $E'$, containing $E''$,
satisfying (T1)-(T2), and such that the degree of $t_i$ is
$2d_i$, $1\leq i\leq p-2$. 

\subsubsection{Solving the Subproblems}
\label{subsubsec:subproblems}

Without loss of generality assume that the $p-1$ cities
connecting the tunnels in $T'$ are
$\{c_1,...,c_{p-1}\}$. Let $C'=\{c_1,...,c_{p-1}\}$, and $C''=C-C'$.
For each $i=1,...,p$, we split the tunnel $t_i$ into two nodes,
$t_i^B$ and $t_i^F$. They are called, respectively, $B$ and $F$ tunnel
entrances.
Define $T'_B=\{t_1^B,...,t_p^B\}$ and $T'_F=\{t_1^F,...,t_p^F\}$.

Let $G'=(C''\cup (T_B'\cup T_F'),E^*)$ be an edge-weighted,
bipartite multigraph with two edges,
each of weight $B(c_j,t_i)$, connecting $c_j$ and $t_i^B$,
$j=p,p+1,...,n$, $i=1,...,p$, and two edges,
each of weight $F(c_j,t_i)$, connecting $c_j$ and $t_i^F$,
$j=p,p+1,...,n$, $i=1,...,p$.

Next, using the notation from the previous subsection,
for each $i=1,...,p$,
let $f_i^B= |E''\cap E[t_i,B]|$ and $f_i^F=|E''\cap E[t_i,F]|$.
(Note that $f_i^B$ ($f_i^F$) is the number of ``back'' (``front'')
type edges in $E''$ that are incident to tunnel $t_i$.)

Following is a formal description of the maximization problem:

$$ g(d_p)= \max (\sum_{i=1}^p \sum_{j=p}^n B(c_j,t_i)x_{i,j}^B+
\sum_{i=1}^p \sum_{j=p}^n F(c_j,t_i)x_{i,j}^F)$$

subject to
$$  \sum_{j=p}^n x_{i,j}^B= d_i-f_i^B,\ i=1,...,p-2,p$$
$$  \sum_{j=p}^n x_{i,j}^F= d_i-f_i^F,\ i=1,...,p-2,p$$
$$\sum_{j=p}^n x_{p-1,j}^B= n-d_p-f_{p-1}^B- \sum_{i=1}^{p-2}d_i,$$
$$\sum_{j=p}^n x_{p-1,j}^F= n-d_p-f_{p-1}^F- \sum_{i=1}^{p-2}d_i,$$
$$  \sum_{i=1}^p (x_{i,j}^B+x_{i,j}^F)=2,\ j=p,...,n,$$
$$  x_{i,j}^B\ge 0,\ i=1,...,p,\ j=p,...,n, $$
$$  x_{i,j}^F \ge 0,\ i=1,...,p,\ j=p,...,n. $$

It is easy to see that for each integer value $d_p\in \{1,2,...,n\}$,
the above problem is an instance of the classical transportation
problem with $n-p+1$ sources (cities) and $2p$ destinations ( $p$
B tunnels and $p$ F tunnels).  Therefore there is an optimal
solution to the linear program for each integer  value of $d_p$ in
which all variables are integer. Moreover, since the number of
destinations is fixed ($p\le k$), when $d_p$ is specified the
dual of this transportation problem can be solved in $O(n)$ time
by the algorithm in Zemel \cite{Zemel}. (See also Megiddo and Tamir
\cite{MeTa93}.)
In particular, $g(d_p)$ can be computed by solving the
dual in $O(n)$ time.
Viewing $d_p$ as a (real) parameter, we note that $g(d_p)$, the
optimal objective value of the above transportation problem,
is a concave function of $d_p$.
To see this, observe that if $d < d'$ are legal values for $d_p$,
and $\bar{x}_d$ and $\bar{x}_{d'}$ are the optimal solutions for
these two values viewed as vectors of variable values, then
$(\bar{x}_d+\bar{x}_{d'})/2$ is a feasible solution for $(d+d')/2$
and so $g((d+d')/2) \geq (g(d)+g(d'))/2$.
Thus,
we can apply a Fibonacci search over the integers $\{1,2,...,n\}$,
or alternatively perform a binary search, at each step checking the values 
at $d-1$, $d$, and $d+1$.
(Actually, $d_p$ is restricted by $1\le d_p
<n-\sum_{i=1}^{p-2}d_i$.)
Specifically, by computing the function $g(d_p)$ at $O(\log n)$
values of $d_p$ we obtain the  (integer)
value of $d_p$ maximizing $g(d_p)$.
Thus, in $O(n \log n)$ time we find the best value of $d_p$.

Therefore, in time $O(n^{2k-3} \cdot n\log n) = O(n^{2k-2}\log n)$ we
can
determine the set of identifiers for the transportation problem that
yields the maximum solution value (including the full degree sequence
for the tunnels).  This solution value will be the length of the
maximum length tour, but because we were solving the dual rather than
primal LP's, we won't yet have the optimal tour itself.  For this,
we need only solve the optimal transportation problem directly in its
primal form.  Since the number of tunnels is bounded, we can do this
in deterministic time $O(n)$ using an algorithm described in an unpublished
paper by Matsui~\cite{Matsui}.  
The basic idea is to apply complementary slackness
properties to the already computed dual solution variables to reduce the
problem to a network flow problem for a bounded number of sinks.
This can then be solved in linear time using an algorithm of
of Gusfield, Martel and
Fernandez-Baca\ \cite{GMF} (see also Ahuja, Orlin, Stein and
Tarjan \cite{AOST}).
Alternatively, one can use the published algorithm
of Tokuyama and Nakano \cite{TN} that requires time $O(n \log^2 n)$ time.
In either case, the time to find this one primal
solution is dominated by the time already spent to
solve all the duals, so the overall running time bound
is that for the latter: $O(n^{2k-2}\log n)$.

\subsection{Further Improvement}
\label{subsec:further}

One way to improve upon the above  bound  is to reduce the running
time to solve an instance of the parametric transportation problem from
$O(n \log n)$ to $O(n)$.  At this point we do not know how to
achieve the linear bound, but we feel that the following
approach might be fruitful.

Consider the above parametric transportation problem, and view the
parameter $d_p$ as an additional (real) variable. The resulting linear
program, which we call the primal, is not a transportation problem.
Nevertheless, the
algorithms in Zemel \cite{Zemel} and Megiddo and Tamir \cite{MeTa93}
can still
solve the dual of this linear program in $O(n)$ time. The
missing ingredient at this point is how to use the dual solution
to obtain (in linear time),
an optimal solution to the primal program. Specifically, we  only need
to know the
optimal (real) value of the primal variable $d_p$, say $d_p^*$.
If $d_p^*$ is known, we  can use the
concavity property of the function $g(d_p)$ to conclude that
the optimal integer value of $d_p$ is attained by rounding $d_p^*$
up or down.
We can then proceed as above.

%In the full paper we show how two additional ideas enable
%us to reduce our running times to $O(n^{2k-2}\log n)$,
%as needed for the proof of Theorem \ref{th:main}.
%The first idea is to view each
%$b$-matching problem as a transportation problem with a bounded number
%of customer locations.
%This latter problem can be solved in linear time by combining
%ideas from \cite{MeTa,GMF,Zemel}.
%The second idea is to exploit the similarities
%between the transportation instances we need to solve.
%Here a standard concavity result implies that one
%dimension of our search over degree sequences can be handled by a binary search.
%In the full paper we also discuss how the constants
%involved in our algorithms grow with $k$.

\section{An $O(n)$ Algorithm}
\label{sec:linear}

In this section, we describe a linear-time algorithm
for determining the length of an optimal tour for
the Maximum TSP under rectilinear distances in the plane.
This amounts to a proof of Theorem~\ref{th:linear}, 
structured into a series of Lemmas and stretching over 
Subsections~\ref{sub:stars} to ~\ref{sub:merge}.
At the end of the section we sketch how this result can be extended to any
4-facet polyhedral metric. Throughout this section we assume without
loss of generality that $n\geq 4$, as otherwise the problem is trivial.
 
\subsection{Stars and Matchings}
\label{sub:stars}
Our construction uses properties of so-called {\em stars};
a star for a given set of vertices $V$ is a minimum
Steiner tree with precisely one Steiner point (the {\em center})
that contains all vertices in $V$ as leafs.
The total length of the edges in a star
is an upper bound on any matching in $V$, since any
edge $(v_i,v_j)$ in the matching can be mapped to a
pair of edges $(v_i,c)$ and $(c, v_j)$ in the star,
and by triangle inequality, $d(v_i,v_j)\leq d((v_i,c)+d(c, v_j)$.
The worst case ratio between the total length of a minimum
 star $\min S(P)$ and a maximum matching $\max M(P)$ plays a crucial role
 in several different types of optimization problems.
 See the paper by Fingerhut, Suri, and Turner~\cite{FiSuTu97}
 for applications in the context of broadband communication networks.
 Also, Tamir and Mitchell~\cite{TaMi98} have used the duality
 between minimum stars and maximum matchings for showing that
 certain matching games have a nonempty core.
 The value of the worst case ratio under the Euclidean metric has
 been determined by Fekete and Meijer; see \cite{FM} for this result
 and several extensions.

In the rest of this section, all distances are 
planar rectilinear distances, unless noted otherwise 
at the end of the section. 
For rectilinear distances, determining the length $\min S(P)$
of a minimum length star
(also known as the rectilinear planar unweighted 1-median
or rectilinear {\em Fermat--Weber problem}) can be done in linear time.
 
\begin{lemm}
\label{le:starcenter}
Suppose we are given 
a given set of $n$ points $P=\{p_1,\ldots, p_n\}$ with $p_i=(x_i,y_i)$,
$1 \leq i \leq n$.
Then under $L_1$ distances
there is an optimal star center $c=(x_c,y_c)$, where $x_c$ is a median
of $\{x_i\mid i=1,\ldots,n\}$, and $y_c$ is a median of 
$\{y_i\mid i=1,\ldots,n\}$. 
\end{lemm}

This is problem 9-2e in \cite{CLRS}.
Recall that if $n = 2m$ is even, then both the $m$'th and $(m+1)$'st
largest items are medians.)
The proof is immediate; for an $O(n)$ running time, use the result
by Blum, Floyd, Pratt, Rivest, and Tarjan~\cite{linear} for computing
a median of $n$ numbers.

It should be noted that for Euclidean distances, the problem of determining
$\min S(P)$ is considerably harder: It was shown by Bajaj~\cite{Baj88}
that the problem of finding an optimal star center for five points in the plane
is in general not solvable by radicals over the field of rationals.
This implies that an algorithm for computing $\min S(P)$
must use stronger tools than 
constructions by straight edge and compass.
 
\begin{figure}[htbp]
   \begin{center}
   \epsfxsize=.45\textwidth
\ \epsfbox{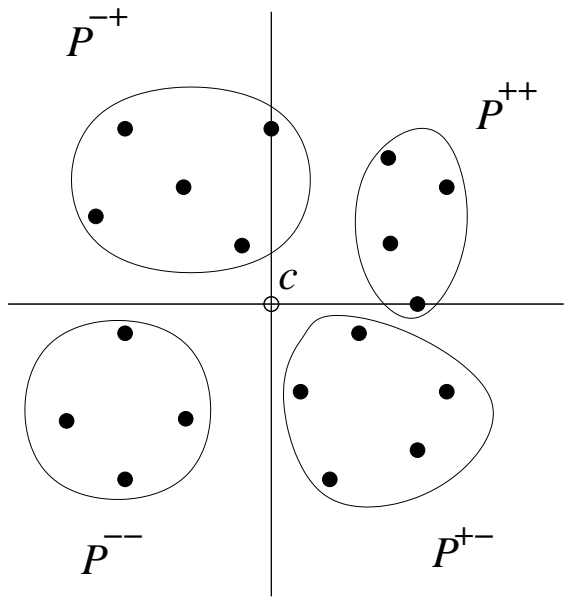}
\caption{The four quadrants and their point sets.}
\label{star}
   \end{center}
\end{figure}
 
The following Theorem~\ref{th:match} appears in the paper by Tamir and
Mitchell~\cite{TaMi98} as Theorem~8; independently, it was noted
by Fekete and Meijer\cite{FM}.

\begin{theo}
\label{th:match}
For $|P|$ even, we have $\max M(P)=\min S(P)$ for rectilinear
distances in the plane.
\end{theo}
 
The basic idea is that the coordinates of an optimal star center
subdivide the plane into four quadrants. If ties are broken in
the right way, the number of points in opposite quadrants
is the same. See Figure~\ref{star}.
Then \begin{equation}
 L_1(v_i, v_j)=L_1(v_i, c)+ L_1(c,v_j) \label{dreieck}
\end{equation}
holds for any edge $(v_i, v_j)$ in the matching, and the theorem follows
easily. \qed

\medskip
Since the results of this paper include the case in which
$|P|$ is odd, we formalize and generalize this observation. 
 
Let $x_c$ and $y_c$ be as in Lemma~\ref{le:starcenter} and define
$P_x^-:=\{p_i\in P\mid x_i<x_c\}$,
$P_x^0:=\{x_i\mid x_i=x_c\}$,
$P_x^+:=\{p_i\in P\mid x_i>x_c\}$,
$P_y^-:=\{p_i\in P\mid y_i<y_c\}$,
$P_y^0:=\{y_i\mid y_i=y_c\}$, and
$P_y^+:=\{p_i\in P\mid y_i>y_c\}$,
{From} the above conditions, it follows that
\begin{eqnarray}
n_x^-:&=&|P_x^-|\leq\frac{n}{2}\label{eins}\\
n_x^+:&=&|P_x^+|\leq\frac{n}{2}\label{zwei}\\
n_y^-:&=&|P_y^-|\leq\frac{n}{2}\label{drei}\\
n_y^+:&=&|P_y^+|\leq\frac{n}{2}.\label{vier}
\end{eqnarray}
Let $n_x^0:=|P_x^0|$,
and $n_y^0:=|P_y^0|$.
By picking any subset of $P_x^{0}$ of size $\lceil\frac{n}{2}\rceil - n_x^-$
and joining it with $P_x^-$, we get a set $P_x^{-/0}$ of size
$\lceil\frac{n}{2}\rceil$; the remaining
$\lfloor\frac{n}{2}\rfloor$ points form the set $P_x^{0/+}$.
Similarly, we get the partition into $P_y^{-/0}$ of size
$\lceil\frac{n}{2}\rceil$, and $P_y^{0/+}$ of size $\lfloor\frac{n}{2}\rfloor$.
Define the following {\em quadrant sets}:
$P^{--}:=P_x^{-/0}\cap P_y^{-/0}$,
$P^{-+}:=P_x^{-/0}\cap P_y^{0/+}$,
$P^{+-}:=P_x^{0/+}\cap P_y^{-/0}$, and
$P^{++}:=P_x^{0/+}\cap P_y^{0/+}$.
The sets $P^{--}$ and $P^{++}$ are {\em opposite quadrant sets},
as are $P^{-+}$ and $P^{+-}$. Two quadrant sets that are
not opposite are called {\em adjacent}.
Finally, let $n^{--}:=|P^{--}|$, etc. 
We get the following conditions:
 
\begin{lemm}
\label{le:cond}
If $n$ is even, then opposite quadrant sets contain the same number
of points, i.\ e.,
\begin{eqnarray}
n^{--}&=&n^{++} \label{eveneins}\\
n^{-+}&=&n^{+-}. \label{evenzwei}
\end{eqnarray}
If $n$ is odd, the numbers of points must satisfy
\begin{eqnarray}
n^{--}&=&n^{++}+1\label{oddeins}\\
n^{-+}&=&n^{+-}.\label{oddzwei}
\end{eqnarray}
\end{lemm}
 
\proof
{From} the definition of the quadrant sets, it follows for even $n$ that
\begin{eqnarray}
n^{--}+n^{-+}=n^{+-}+n^{++}\label{evendrei}\\
n^{--}+n^{+-}=n^{-+}+n^{++}.\label{evenvier}
\end{eqnarray}
{From} this the claims (\ref{eveneins}) and (\ref{evenzwei}) follow easily, 
as was noted in ~\cite{TaMi98} and \cite{FM}.
 
For the odd case, the definition of the quadrant sets
yields the conditions
\begin{eqnarray}
n^{--}+n^{-+}&=&n^{+-}+n^{++}+1\label{odddrei}\\
n^{--}+n^{+-}&=&n^{-+}+n^{++}+1.\label{oddvier}
\end{eqnarray}
This implies (\ref{oddeins}) and (\ref{oddzwei}).
\qed

\medskip
In the following, we will use Lemma~\ref{le:cond} to derive
first an optimal 2-factor, consisting of at most two subtours,
and then argue how these subtours can be merged optimally.
 
\subsection{2-Factors and Trivial Tours}
\label{sub:2fact}
A 2-factor for a set of vertices
is a multi-set of edges that covers each vertex exactly twice.
Since any tour is a 2-factor, a maximal length 2-factor
is an upper bound for the length of a tour.
Using the triangle inequality, it is straightforward to
see that twice the length of a star is an upper bound
for the length of any 2-factor, even when the star is centered
at one of the given vertices. 
Achieving tightness for this bound is the main stepping
stone for our algorithm.
Based on the results of the preceding section, we prove the
following three lemmas. We start with the easiest case:
 
\begin{lemm}
\label{trivial}
%Suppose $n$ is even or $n$ is odd and $|P_x^{0}\cup P_y^{0}|= 1$.
If two of the quadrant sets are empty,
then there is a feasible tour of length $2\min S(P)$, which is
optimal.
\end{lemm}
 
\proof
Note that the conclusion will follow
if we can construct a tour in which all edges satisfy property (\ref{dreieck})
above.  If two of the quadrant sets are empty, it follows from 
Lemma~\ref{le:cond} that these must be opposite.
Without loss of generality, let us assume that they are
$P^{-,-}$ and $P^{+,+}$. 
For the other two sets, any edge $(v_i,v_j)$ between opposite
quadrant sets satisfies property (\ref{dreieck}).
If the number of points in the two opposite
quadrant sets is the same, we can get a tour by jumping back and forth
while there are unvisited points in these quadrant sets.
If $n$ is odd and $p_*=(x_c, y_c)\in P^{--}$, then $p_*$ can be
inserted into any of these edges, while (\ref{dreieck})
will still apply to all edges.
If $n$ is odd and $p_*=(x_c, y_c)\not\in P^{--}$, then 
$P^{--}$ must contain two points, one with $x_c$ as its $x$-coordinate
and one with $y_c$ as its $y$-coordinate.  The tour connects these
two together and then jumps back and forth between quadrants.
All edge lengths again satisfy (\ref{dreieck}).
\qed

\begin{lemm}
\label{odd}
Suppose no quadrant set is empty, 
$n$ is odd and $|P_x^{0}\cup P_y^{0}|> 1$.
Then there is a feasible tour of length $2\min S(P)$, which is
optimal.
\end{lemm}
 
By conditions~\ref{eins} and~\ref{drei} and because
$x_c$ and $y_c$ are both coordinates of points, we know
that $P_x^0\cap P_x^{-/0}$ and $P_y^0\cap P_y^{-/0}$
each must contain a point. 
As $|P_x^{0}\cup P_y^{0}|> 1$,
we can consider two points $p_a\in P_x^{0}$, $p_b\in P_y^{0}$
with $a\neq b$. Let us assume without loss
of generality that when we constructed $P_x^{-/0}$ and $P_y^{-/0}$
we assigned $p_a$ to the former and $p_b$ to the latter.

We distinguish the following cases -- see Figure~\ref{facten}:
 
\begin{figure}[htbp]
   \begin{center}
   \epsfxsize=.95\textwidth
\ \epsfbox{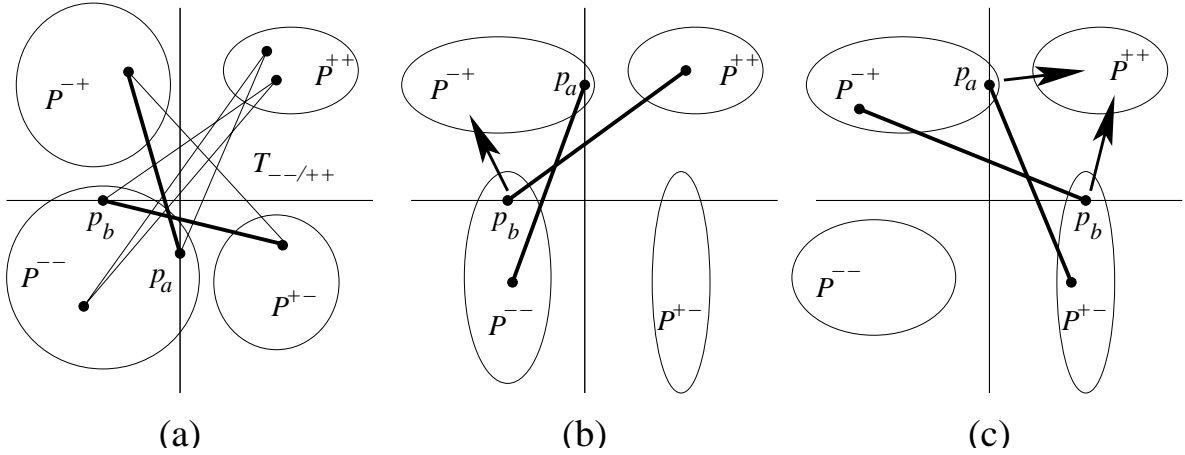}
\caption{Getting optimal 2-factors.}
\label{facten}
   \end{center}
\end{figure}
 
\medskip
{\boldmath $(a)\  p_a, p_b\in P^{--}$:}
 
By connecting $p_a$ with a point in $P^{-+}$,
and $p_b$ with a point in $P^{+-}$, and otherwise
jumping back and forth between
opposite quadrant sets, we get a tour that satisfies (\ref{dreieck})
for any edge.
 
\medskip
{\boldmath $(b1)\ p_a \not\in P^{--}, p_b\in P^{--}$:}
 
In this case, $p_a\in P^{-+}$.
By changing the membership of $p_b$ from $P^{--}$ to $P^{-+}$, we get
$|P^{--}|=|P^{++}|$ and $|P^{-+}|=|P^{+-}|+1$.
Then a tour for the modified $P^{-+}$ and $P^{+-}$
can be obtained as in case (a).
 
\medskip
{\boldmath $(b2)\ p_a \in P^{--}, p_b\not\in P^{--}$:}
 
This is treated in the same way as case (b1).
 
\medskip
{\boldmath $(c)\ p_a, p_b \not\in P^{--}$:}
 
In this case, $p_a\in P^{-+}$ and $p_b\in P^{+-}$.
By changing the membership of $p_a$ from $P^{-+}$ to $P^{++}$
and the membership of $p_b$ from $P^{+-}$ to $P^{++}$, we get
$|P^{++}|=|P^{--}|+1$ and $|P^{-+}|=|P^{+-}|$,
so we can get tours as in case (a).
\qed

\medskip 
In the remaining cases we may no longer be
able to get a tour that meets the $2\min S(P)$ upper bound, but the
next Lemma says that we can construct two disjoint tours whose total
length meets this bound.  We will subsequently show how these tours
can be combined with only a small decrease in total length
to obtain a single optimal tour.
 
\begin{lemm}
\label{twofact}
Suppose no quadrant is empty and either
$n$ is even or $n$ is odd and $|P_x^{0}\cup P_y^{0}|= 1$.
Then there is a tour $T_{--/++}$
of the points in $P^{--}\cup P^{++}$, and a tour $T_{-+/+-}$
of the points in $P^{-+}\cup P^{+-}$, such that
$\ell(T_{--/++})+\ell(T_{-+/+-})=2\min S(P)$.
\end{lemm}
 
\proof
If $n$ is even, we can argue like
in the proof of Lemma~\ref{trivial}:
We get two subtours, one covering each
pair of opposite quadrant sets.
 
If $n$ is odd and
there is only one point $p_*$ in $P_x^{0}\cup P_y^{0}$,
the case reduces to $n$ even,
since $p_*=(c_x, c_y)\in P^{--}$, and $p_*$ can be
inserted into any tour of $P^{--}\setminus \{p_*\}$ and $P^{++}$
while still guaranteeing (\ref{dreieck}) for any tour edge.
\qed

\subsection{How to Merge 2-Factors}
\label{sub:merge}
Suppose that no quadrant set is empty and that
we have a pair of subtours, whose
total length matches the $2\min S(P)$ upper bound on optimal tour length,
as in Lemma~\ref{twofact}. 
Now we shall show how the upper bound has to be adjusted if we are
to restrict ourselves to connected tours, and how the adjusted bound
can be met.

We start with the easier case of $n$ odd and the median being part of the
point set, before dealing with
the more complicated
case of even $n$. 

\subsubsection{Odd $n$}
Let $n$ be odd and 
$|P_x^{0}\cup P_y^{0}|= 1$, which means that $p_*=(x_c,y_c)$ is in $P$. 

\begin{lemm}
\label{le:adjacent}
Let $n$ be odd, 
$|P_x^{0}\cup P_y^{0}|= 1$, and all quadrant sets be nonempty.
Then any tour of $P$ contains an edge that connects two adjacent 
quadrant sets and is not incident on $p_*$.
\end{lemm}

\proof
Any tour $T$ of $P$ induces a tour $T'$ on the set $P\setminus\{p_*\}$;
$T'$ must contain at least two different edges $e_1$ and $e_2$
that connect adjacent quadrant sets,
i.\ e., that connect $S_1=(P^{--}\setminus \{p_*\}\cup P^{++})$ to
$S_2=(P^{-+}\cup P^{+-})=P\setminus S_1$. One of these two edges
must also be part of $T$, and the claim follows.
\qed

\begin{lemm}
\label{oneedge}
Let $e_1=(p_1, p_2)$ be an edge connecting two horizontally
(or vertically) adjacent quadrant sets.
Let $p_i=(x_i, y_i)$, and define $z:=\min\{|y_c-y_1|, |y_c-y_2|\}$
(or $z:=\min\{|x_c-x_1|, |x_c-x_2|\}$).
Then any tour containing $e_1$ has length at most $2\min S(P)-2z$.
\end{lemm}
 
\proof
In either case,
we have $L_1(p_1, p_2)=L_1(p_1,c)+L_1(c,p_2)-2z$, and
the claim follows.
\qed

\medskip 
By considering all possible edges $(p_1,p_2)$ connecting adjacent
quadrant sets, we get an adjusted upper bound
on the tour length. Note that only the smaller distance from
a median line matters for this bound.
More formally, let
$Z_1=\min\{|y_c-y_i|\mid p_i\in P\setminus\{p_*\}\}$,
and $Z_2=\min\{|x_c-x_i|\mid p_i\in P\setminus\{p_*\}\}$,
and let $Z_*=\min\{Z_1,Z_2\}$.

\begin{lemm}
\label{oddadjust}
Let $n$ be odd, 
$|P_x^{0}\cup P_y^{0}|= 1$, and all quadrant sets be nonempty.
Then an optimal tour of $P$ has length
$\min S(P)-2 Z_*$, and such a tour can be found
in linear time.
\end{lemm}

\begin{figure}[htbp]
   \begin{center}
   \epsfxsize=.40\textwidth
\ \epsfbox{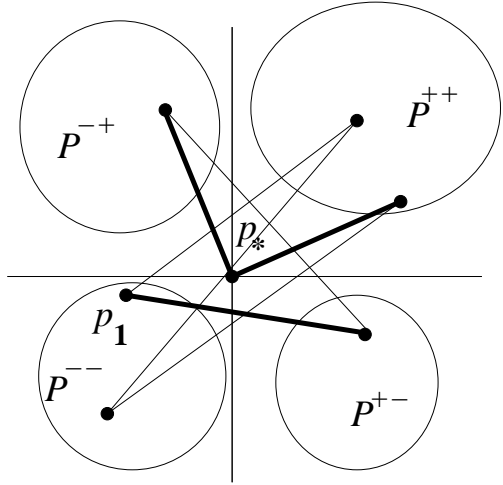}
\caption{Getting an optimal tour for odd $n$, when the median is in $P$.}
\label{oddbound}
   \end{center}
\end{figure}

\proof
By Lemma~\ref{le:adjacent} and Lemma~\ref{oneedge},
$\min S(P)-2 Z_*$ is a valid upper bound on the tour length.
It follows from the discussion of $S(P)$ and the definition
of $Z_*$ that the bound can be computed in linear time.
Finally suppose for example that $Z_* = Z_1$ and that
$p_1 \in P^{--}$ is a point for which the value of $Z_1$ is met.
Connect $p_1$ to a vertex in $P^{+-}$, and 
$p_*$ to vertics in $P^{++}$ and $P^{-+}$.
As shown in Figure~\ref{oddbound}, it is 
straightforward to add only edges between opposite
quadrants in order to get a tour of the required length.
Other cases are handled analogously.
\qed

\subsubsection{Even $n$}
The proof proceeds similarly to the
case for odd $n$ but requires a more involved combinatorial analysis.
Let us say that a pair of edges $e_1=(v_1,v_2)$ and $e_2=(v_3,v_4)$ is
a {\em quadrant matching of the first type} if
$v_1 \in P^{--}$, $v_2 \in P^{-+}$, $v_3 \in P^{+-}$, $v_4 \in P^{++}$ or if
$v_1 \in P^{--}$, $v_2 \in P^{+-}$, $v_3 \in P^{-+}$, $v_4 \in P^{++}$.
See Figure 4.  We will call $e_1$ and $e_2$ a {\em quadrant matching
of the second type} if one edge joins adjacent quadrants and the other
lies within a third quadrant.
 
\begin{figure}[htb]
   \begin{center}
   \epsfxsize=.55\textwidth
\ \epsfbox{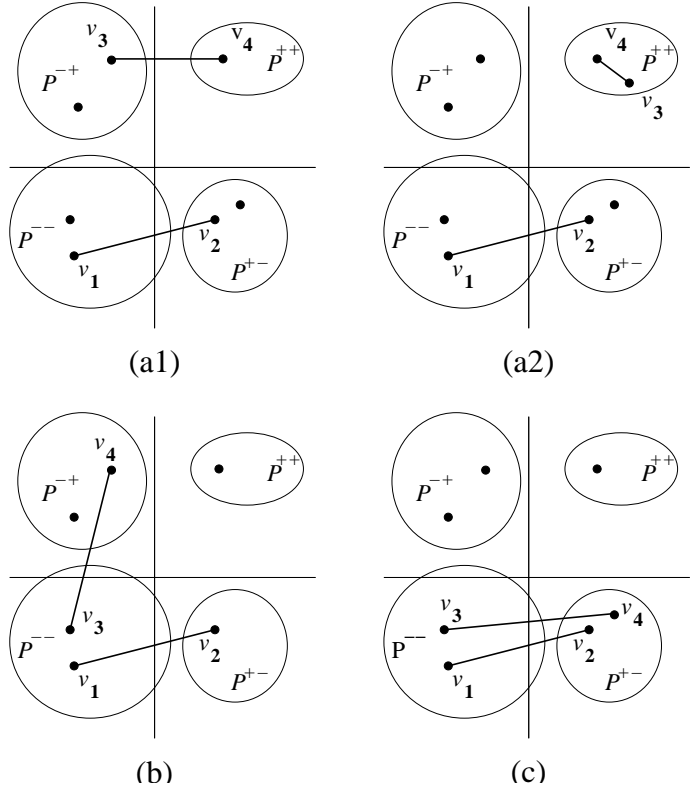}
\caption{(a1+2) Quadrant matchings of the first and second type. 
(b+c) Edges connecting adjacent quadrant sets.}
\label{pair}
   \end{center}
\end{figure}
 
\begin{lemm}
\label{quad}
Let $n$ be even and all quadrant sets be nonempty.
Then any tour of $P$ contains a quadrant matching
of either the first or the second type.
\end{lemm}
 
\proof
As in the proof of Lemma~\ref{le:adjacent},
any tour of $P$ must contain at least two different edges $e_1=(v_1,v_2)$ 
and $e_2=(v_3,v_4)$
that connect adjacent quadrant sets,
i.\ e., that connect $S_1=(P^{--}\cup P^{++})$ to
$S_2=(P^{-+}\cup P^{+-})=P\setminus S_1$.

We consider the following cases; in all cases, we name particular
choices of quadrants for clearer notation and better reference to 
Figure~\ref{pair}. All other choices are completely analogous.
Moreover, all arguments remain valid if two of the named vertices
coincide.
 
\medskip
{\boldmath $(a)\ \ (v_1, v_2), (v_3, v_4)$} {\bf form a quadrant matching
of either type}:
 
In this case, there is nothing to prove.
 
\medskip
{\boldmath $(b)\ \ (v_1, v_2), (v_3, v_4)$} {\bf connect one quadrant
with both adjacent quadrants:}

Suppose without loss of generality
$v_1, v_3\in P^{--}$, $v_2\in P^{+-}, v_4\in P^{-+}$, as shown in the figure.
Since two edges adjacent to vertices in $P^{--}$
are already given,
there can be at most $2 n^{--} - 2= 2 n^{++} - 2$ edges between
$P^{--}$ and $P^{++}$, so there must be an edge connecting two
vertices in $P^{++}$, or two edges connecting
$P^{++}$ to adjacent quadrant sets.
In the first case we have a quadrant
matching of the second type and in the second case we have a quadrant
matching of the first type, so the claim follows.

\medskip
{\boldmath $(c)\ \ (v_1, v_2), (v_3, v_4)$} {\bf connect the same
pair of adjacent quadrants:}

Suppose without loss of generality $v_1, v_3\in P^{--}$, $v_2, v_4\in P^{+-}$.
Since two edges adjacent to vertices in $P^{+-}$ are already given,
there can be at most $2 n^{+-} - 2= 2 n^{-+} - 2$ edges between
$P^{-+}$ and $P^{+-}$, so there must be an edge connecting two
vertices in $P^{-+}$ (and we are done), or there must be two edges between
$P^{-+}$ and adjacent quadrant sets, either 
yielding a quadrant matching of the first type (and we are done), 
or reducing this case to case (b).
\qed

\medskip 
Now we can give an upper bound on the length of an optimal tour with a
given pair of edges.
 
\begin{lemm}
\label{twoedges}
Let $e_1=(p_1, p_2)$ be an edge connecting two
adjacent quadrant sets, say,
$P^{--}$ and $P^{+-}$.
Let $e_2=(p_3,p_4)$ be an edge forming a quadrant matching with
$e_1$.
Let $p_i=(x_i, y_i)$, and define $z_1:=\min\{(y_c-y_1), (y_c-y_2)\}$,
$z_2:=\min\{(y_3-y_c), (y_4-y_c)\}$.
 
Then any tour
containing $e_1$ and $e_2$ has length at most $2\min S(P)-2z_1 - 2z_2$.
\end{lemm}
 
\proof
Since $L_1(p_1, p_2)=L_1(p_1,c)+L_1(c,p_2)-2z_1$, and
$L_1(p_3, p_4)\leq L_1(p_3,c)+L_1(c,p_4)-2z_2$, the claim follows.
\qed

\medskip 
By considering all pairs of edges, we get an adjusted upper bound
on the tour length. For this purpose, let
$Z_1=\min\{|y_c-y_i|\mid p_i\in P^{--}\cup P^{+-}\}$,
and let
$Z_2=\min\{|y_i-y_c|\mid p_i\in P^{-+}\cup P^{++}\}$.
Similarly, let
$Z_3=\min\{|x_c-x_i|\mid p_i\in P^{--}\cup P^{-+}\}$, and
let
$Z_4=\min\{|x_i-x_c|\mid p_i\in P^{+-}\cup P^{++}\}$.
Finally, let $Z_*=\min\{Z_1+Z_2, Z_3+Z_4\}$.
 
\begin{lemm}
\label{adjust}
Let $n$ be even, and all quadrant sets be nonempty. 
Then an optimal tour of $P$ has length
$\min S(P)-2 Z_*$, and such a tour can be found
in linear time.
\end{lemm}
 
\begin{figure}[hbtp]
   \begin{center}
   \epsfxsize=.40\textwidth
\ \epsfbox{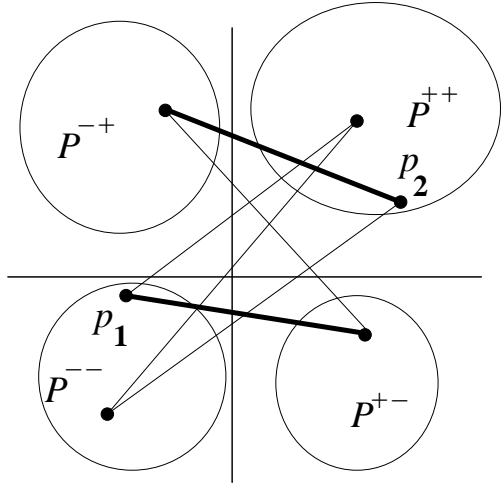}
\caption{Getting an optimal tour for even $n$.}
\label{bound}
   \end{center}
\end{figure}
 
\proof
It follows immediately from Lemmas~\ref{quad},~\ref{twoedges}
and the definition of $Z_*$ that $\min S(P)-2 Z_*$
is a valid upper bound that can be computed in linear time.
To see that there is a tour of this length, consider a pair of vertices 
where the
value $Z_*$ is met. Without loss of generality, let this be for
$p_1\in P^{--}$ and $p_2\in P^{++}$. Connect $p_1$ to any vertex
in $P^{+-}$, and $p_2$ to any vertex in $P^{-+}$. Now
it is easy to see that using only edges connecting opposite
quadrant sets, we can get a tour. See Figure~\ref{bound}.
\qed

This concludes the proof of Theorem~\ref{th:linear}. 

\subsection{An extension}
\begin{figure}[htbp]
   \begin{center}
   \epsfxsize=.9\textwidth
\ \epsfbox{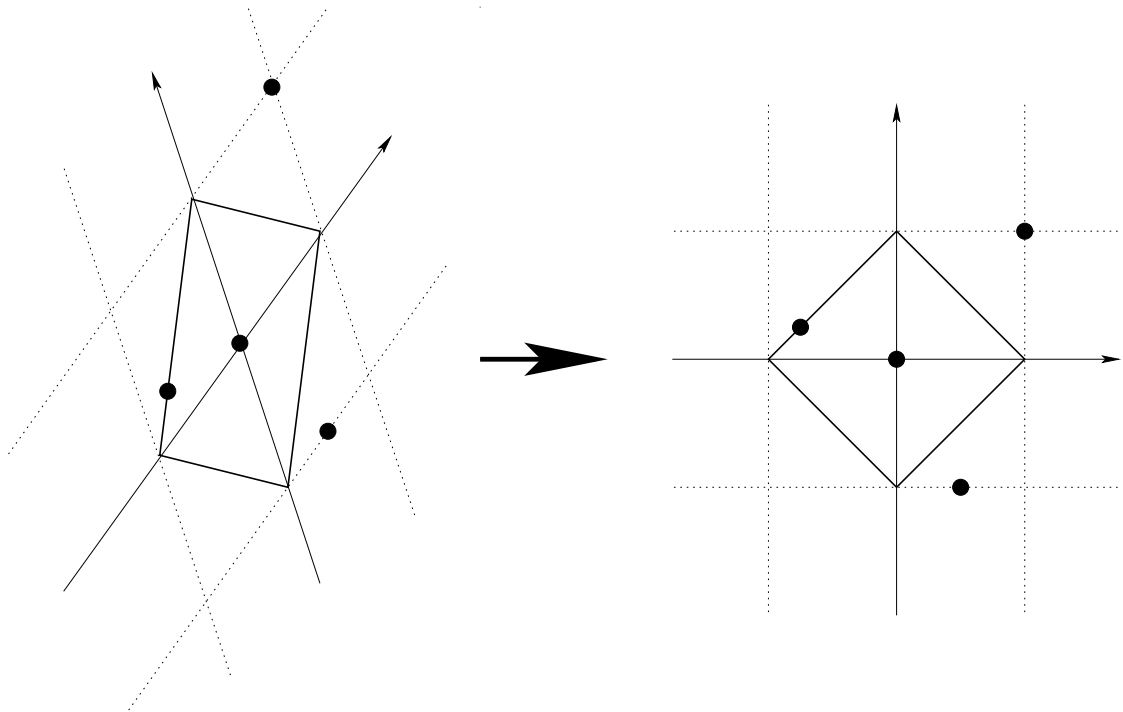}
\caption{Transforming an arbitrary 4-facet norm into the $L_1$ norm.}
\label{4facet}
   \end{center}
\end{figure}

Using a rotation by $\pi/4$, it is easy to transform $L_\infty$ distances
to $L_1$ distances, so the theorem remains valid for this case.
Moreover, any 4-facet polyhedral metric can be transformed into
the $L_1$ metric 
with an appropriate coordinate
transformation, turning the unit ball into a square.
See Figure~\ref{4facet} for an illustration.
All arguments described in the preceding section can still
be applied for these transformed coordinates, so the following
generalization holds: 
 
\begin{theo}
\label{th:4-face}
For any 4-facet polyhedral metric in the plane, a tour of maximum
possible length can be constructed in linear time.
\end{theo}

We conclude this section by noting that
Theorem~\ref{th:match} does not appear to extend to two-dimensional metrics
with more than four facets, the two-dimensional Euclidean metric,
or the rectilinear metric in higher dimensions. 

To see that there is no easy generalization even for $L_1$
distances in higher dimensions,
note that the partition into orthants by an optimal star
center may not induce a ``balanced'' partition of the point
set, such that we have subsets of equal size in opposite orthants.
 
\begin{exam}
Consider $P$ with $\frac{n-1}{4}$ points in each of the orthants
$\{q=(x,y,z)\mid x>0, y>0, z>0\}$,
$\{q=(x,y,z)\mid x<0, y<0, z>0\}$,
$\{q=(x,y,z)\mid x<0, y>0, z<0\}$,
$\{q=(x,y,z)\mid x>0, y<0, z<0\}$,
plus the point $(0,0,0)$. 
Then $(0,0,0)$ is the unique optimal star center.
No connection of points in different orthants keeps the triangle
inequality tight.
\end{exam}
 
That there is a fundamental
difference between the Euclidean and rectilinear metrics for the
plane is clear from the following. 
 
\begin{coro}
\label{char}
For any set of $n$ points in the plane, there are
$\Omega\left((\frac{n}{4}!)^4\right)$ many tours which are optimal
for the Maximum TSP under rectilinear distances.
If the distances are Euclidean, there may only be one optimal tour.
\end{coro}
 
\proof
Any tour that can be constructed
as in Lemma~\ref{trivial},
Lemma~\ref{oddadjust}, or Lemma~\ref{adjust} is optimal, 
so we can choose an arbitrary
permutation for each quadrant set. This yields the above lower
bound on the number of optimal tours. 
Conversely, we see that any optimal tour must have the structure
described in the Lemmas.

To see that there may
only be one optimal tour for Euclidean distances, consider
a set of $n=2k+1$ points that are evenly distributed
around a unit circle.
\qed

\section{NP-Hardness Results}
\label{sec:npc}

Now we proceed to show that changing from polyhedral to Euclidean
distances, or to distances in spaces of unbounded dimension, changes the
problem complexity from polynomial (or even linear) to NP-hard. 
This dramatic effect illustrates that failing to model geometric
instances of optimization problems beyond their
combinatorial graph structure can miss out on important 
differences in problem complexity.

\subsection{Euclidean Distances in 3D}
\label{subsec:L23D}
In this section, we establish the NP-hardness of the Maximum
TSP under Euclidean distances in $\Rd$.
The proof gives a reduction of the well-known
problem {\em Hamilton Cycle in Grid Graphs}, which was shown to be
NP-complete by
Itai, Papadimitriou, and Swarcfiter~\cite{ItPaSw}.
A grid graph $G$ is given by a finite set of vertices
$V=\{v_1\,v_2,\ldots,v_n\}$, with each vertex $v_i$
represented by a grid point $(x_i, y_i)\in Z^2$;
for easier notation, we write $v_i=(x_i,y_i)$. 
Two vertices
$v_i$ and $v_j$ in $G$ are adjacent if and only if they are at distance
1, i.\,e., if $(x_i-x_j)^2+(y_i-y_j)^2=1$.
Without loss of generality, we may assume that $G$ is connected,
and that $n$ is sufficiently large.
Note that any grid graph is bipartite: vertices
$v_i$ with $x_i+y_i$ even can only be adjacent to
vertices $v_j$ with $x_j+y_j$ odd, and vice versa.
In the following, we will denote this partition by
$V=V_e\dot\cup V_o$, where $V_e$ is the set of 
vertices with even coordinate sum, while $V_o$ is the set of vertices
with odd coordinate sum.
 
The basic idea of the proof is to embed any grid graph $G$
into the surface of a sphere in $\Rr$, such that edges in
the grid graph correspond to longest distances within the point
set. This can be achieved
by representing the vertices in $V_e$ by points that are
relatively close to each other around
a position $(a,b,c)$ on the sphere,
and the vertices in $V_o$ by points
close to each other at a position on the sphere
that is roughly opposite (i.\,e., antipodal) to $(a,b,c)$; for simplicity
of description by spherical coordinates, we will
use positions that are close to the equator. Locally, the mapping
of the two point sets onto the sphere is an approximation
of the relative position of vertices in the grid graph. Since
adjacent vertices in a grid graph have different parity,
unit edges in the grid graph representation correspond to edges connecting
points that are almost at opposite positions on the sphere,
and vice versa.
 
In the following,
the technical details are described. For simplicity, we use
spherical coordinates and multiples of $\pi$. However,
it will become clear from our discussion that we only require
computations of bounded accuracy. It is straightforward
to use only Cartesian coordinates that can be obtained
by polynomial time approximation within the desired overall error bound of
$O\left(n^{-8}\right)$ for the length of an edge.
 
Represent each vertex $v_i$ by a point $S(v_i)$
on the unit sphere, described by
spherical coordinates $(r,\phi,\theta)$, which translate
into Cartesian coordinates by
$x=r \cos\phi \cos\theta$,
$y=r \sin\phi \cos\theta$,
$z=r \sin\theta$.
Note that, as in standard geographic coordinates,
the ``equator'' of the sphere is given
by $\theta=0$; the angle $\theta$ describes the ``latitude''
of a point, while $\phi$ describes the ``longitude''.
Since we will only consider points with $r=1$,
we will simply write $(\phi,\theta)$ in spherical coordinates,
but $(x,y,z)$ in Cartesian coordinates.
Let $\psi=\frac{2\pi}{n^3}$.
Now any vertex $v_i\in V_e$ is represented by
a point $S(v_i)=(x_i\psi, y_i\psi)$.
Any vertex $v_i\in V_o$ is represented by
a point $S(v_i)=(\pi + x_i\psi, -y_i\psi)$.
 
\begin{lemm}
\label{long}
There is a small constant $\eps_n=O\left(n^{-8}\right)$, which can be computed
in polynomial time, such that
for the three-dimensional Euclidean distance $L_2$ between
two points $S(v_i)$ and $S(v_j)$, the relation
$L_2\left(S(v_i),S(v_j)\right)\geq 2-\frac{\psi^2}{4}-\eps_n$ holds
if and only if $v_i$ and $v_j$ are adjacent in $G$.
In that case, 
$L_2\left(S(v_i),S(v_j)\right)\leq 2-\frac{\psi^2}{4}+\eps_n$.
If $v_i$ and $v_j$ are not adjacent in $G$, then
$L_2\left(S(v_i),S(v_j)\right)\leq 2-\sqrt{5}\frac{\psi^2}{4}+\eps_n$. 
\end{lemm}
 
\proof
Since the diameter of the grid graph cannot exceed $n$,
it is easy to see that we have 
$L_2\left(S(v_i),S(v_j)\right)\leq n\psi = O\left(n^{-2}\right)$
whenever $v_i$ and $v_j$ have the same parity.
Therefore consider
$v_i\in V_e$ and $v_j\in V_o$.
Then
\begin{eqnarray*}
&&\left[L_2\left(S(v_i),S(v_j)\right)\right]^2\\
&=&
\left[L_2\left(\left(\cos(x_i\psi)\cos(y_i\psi), \sin(x_i\psi)\cos(y_i\psi), \sin(y_i\psi)\right), \right. \right.\\
&&\left.\left.\left(\cos(\pi+x_j\psi)\cos(-y_j\psi), \sin(\pi+x_j\psi)\cos(-y_j\psi), \sin(-y_j\psi)\right)\right)\right]^2\\
&=&
\left[\cos(x_i\psi)\cos(y_i\psi) + \cos(x_j\psi)\cos(y_j\psi)\right]^2
%\\ &&
+ \left[\sin(x_i\psi)\cos(y_i\psi) + \sin(x_j\psi)\cos(y_j\psi)\right]^2
\\ &&
+ \left[\sin(y_i\psi) + \sin(y_j\psi)\right]^2                             \\
&=&
\left[ \left(1-\frac{(x_i\psi)^2}{2}+O\left((x_i\psi)^4\right)\right)
\left(1-\frac{(y_i\psi)^2}{2}+O\left((y_i\psi)^4\right)\right) \right.
\\ &&
+ \left. \left(1-\frac{(x_j\psi)^2}{2}+O\left((x_j\psi)^4\right)\right)
\left(1-\frac{(y_j\psi)^2}{2}+O\left((y_j\psi)^4\right)\right) \right]^2
\\ &+&
\left[ \left(x_i\psi - O\left((x_i\psi)^3\right)\right)
\left(1-\frac{(y_i\psi)^2}{2}+O\left((y_i\psi)^4\right)\right)\right.
\\ && 
+ \left. \left(x_j\psi - O\left((x_j\psi)^3\right)\right)
\left(1-\frac{(y_j\psi)^2}{2}+O\left((y_j\psi)^4\right)\right) \right]^2
\\ &+&
\left[y_i\psi - O\left((y_i\psi)^3\right) + y_j\psi - O\left((y_j\psi)^3\right)\right]^2 \\
&=&
\left[ 2 - \frac{(x_i\psi)^2}{2} -\frac{(y_i\psi)^2}{2}
-\frac{(x_j\psi)^2}{2} -\frac{(y_j\psi)^2}{2} +O\left(n^{-8}\right)\right]^2 \\
&& + \left[ x_i\psi + x_j\psi + O\left(n^{-6}\right)\right]^2 +
\\ &&+
\left[ y_i\psi + y_j\psi + O\left(n^{-6}\right)\right]^2 +
\\ &=&
\left[ 4 - 2{(x_i\psi)^2} -2{(y_i\psi)^2}
-2 {(x_j\psi)^2} - 2 {(y_j\psi)^2} +O\left(n^{-8}\right)\right] \\
&& + \left[ (x_i\psi)^2 + (x_j\psi)^2 + 2x_ix_j\psi^2 + O\left(n^{-8}\right)\right] +
 \left[ (y_i\psi)^2 + (y_j\psi)^2 + 2 y_iy_j\psi^2 + O\left(n^{-8}\right)\right]
 \\
&=&
4-(x_i-x_j)^2\psi^2-(y_i-y_j)^2\psi^2+O\left(n^{-8}\right).
\end{eqnarray*}

Since $v_i$ and $v_j$ have different parity, we have
$(x_i-x_j)^2\psi^2+(y_i-y_j)^2\psi^2=\psi^2$
if $v_i$ and $v_j$ are adjacent in $G$,
and $(x_i-x_j)^2\psi^2+(y_i-y_j)^2\psi^2\geq 5\psi^2$
if $v_i$ and $v_j$ are not adjacent in $G$,
so the claim follows.
\qed

\medskip 
{From} Lemma~\ref{long}, it is straightforward to conclude
that there is a tour of length at least $2n-n\frac{\psi^2}{4}-n\eps_n$,
if and only if the grid graph $G$ is Hamiltonian.
 
By setting additional coordinates equal to zero, we conclude the proof
of Theorem~\ref{th:nphard} for arbitrary dimensions $d\geq 3$.

\subsection{Higher-Dimensional Implications}
\label{subsec:high}
\nopagebreak
There are various implications of Theorem~\ref{th:nphard} to higher dimensions.
It is not hard to generalize the construction and
the proof of Lemma~\ref{long}
to the case of $L_p$ norms, as long as $p\not\in\{1,\infty\}$:
Instead of using a unit sphere for the embedding, use an
$L_p$ ball of dimension 3, which has a smooth surface
whenever $p\neq 1,\infty$. The error bounds can be worked out
in an analogous way.

Combined with Theorem~\ref{th:main}, we note:

\begin{coro}
\label{Lp}
Provided that P$\neq$NP, the Maximum TSP under an $L_p$ norm
in $\Rd$ with $d\geq 3$ fixed
is solvable in polynomial time, if and only if $p\in\{1,\infty\}$.
\end{coro}

There is a close
connection between the Euclidean norm and polyhedral norms
when the number of facets $k$ is not fixed,
as was pointed out by Joe Mitchell~\cite{joe.priv}:
Since we only need to consider $O\left(n^2\right)$ directions for
connections between points, we can replace the Euclidean distances $L_2$
by a polyhedral norm with $O\left(n^2\right)$ facets.
 
\begin{coro}
\label{poly}
The Maximum TSP under a polyhedral norm having a unit ball with $k$ facets
in $\Rd$ is an NP-hard problem, if $d\geq 3$ and $k$ is part of the input.
\end{coro}
 
Another easy consequence concerns the {\em Maximum Scatter TSP},
which was first considered by Arkin, Chiang, Mitchell,
Skiena, and Yang~\cite{ACMSY}. In this problem, the objective
is to find a tour that maximizes the length of the shortest edge.
Arkin et al.~gave an NP-hardness proof for the general case
and a 2-approximation that uses only triangle inequality.
The complexity for geometric instances was left as an open problem.
Using the above construction and Lemma~\ref{long}, we get:
 
\begin{coro}
\label{scatter}
The Maximum Scatter TSP
under Euclidean distances in $\Rd$ is an NP-hard
problem if $d\geq 3$.
\end{coro}
 
Finally, it is straightforward with the above construction
to show the following:
 
\begin{coro}
\label{sphere}
The Maximum TSP and the Maximum Scatter TSP under shortest distances
on the $(d-1)$-dimensional surface of the $d$-dimensional unit sphere
$S^{d-1}$ are NP-hard for $d\geq 3$.
\end{coro}
 
It was also noted by Joe Mitchell that another corollary can be derived
by using an approximation with $O\left(n^2\right)$ facets:
 
\begin{coro}
\label{polytope}
The Maximum TSP and the Maximum Scatter TSP on the
$(d-1)$-dimensional surface of a $d$-dimensional convex
polytope with an unbounded number of facets
under geodesic distances are NP-hard for $d\geq 3$.
\end{coro}
 
Another set of questions concerns the complexity of the Maximum TSP when 
$d$ is {\em not\/} fixed.
It is relatively easy to show that the problem is NP-hard for $L_p$ 
norms:

\begin{theo}
\label{th:high.d.Lp}
The Maximum TSP and the Maximum Scatter TSP are NP-hard for
points in $d$-dimensional space under $L_p$ norms,
$1<p\leq\infty$, when $d$
is part of the input.
\end{theo}

\proof
We use a transformation from
the Hamiltonian Circuit problem for simple cubic graphs $G=(V,E)$,
which was shown to be NP-complete by Garey, Johnson, and Tarjan~\cite{GJT}.
We use a separate
dimension for each edge $e\in E$, and a point $p_i$ for each vertex 
$v_i\in V$: For edge $e=(v_i,v_j)$, choose the $x_e$-coordinate of one
of the points 
(say, $p_i$) to be $1$, and the $x_e$-coordinate of the other point 
(say, $p_j$) to be $-1$. All other $x_e$-coordinates are chosen to be
$0$. This means that the $L_p$-distance between points representing
adjacent vertices in $G$ is the ``long'' value $L_{\max}:=\sqrt[p]{2^p+4}$
(2 for $L_\infty$), 
while nonadjacent vertices
get the ``short'' distance $L_{\min}:=\sqrt[p]{6}$ (1 for $L_\infty$). 
For fixed
$p>1$, it is straightforward to compute a critical value $L < L_{max}$
in polynomial time such that there is a tour of length $nL$ or greater
if and only if there is a Hamiltonian cycle in $G$.
\qed

\medskip
This leaves open the case of the
$L_1$ metric when $d$ is not fixed, although we conjecture that the $L_1$
case is NP-complete as well.  Also open is the question of whether there
might be a PTAS for any such norm when $d$ is not fixed.
Trevisan \cite{Trev} has shown that the Minimum TSP is Max-SNP-hard for any
such norm, and so cannot have such PTAS's unless P = NP.
We can obtain a similar result for the Maximum TSP under $L_{\infty}$ by
modifying our NP-hardness transformation
to prove Max-SNP-hardness
and hence by~\cite{pcp} the existence of an $\epsilon$ such that no polynomial
time approximation algorithm can guarantee a solution within a factor
of $1+\epsilon$ of optimal unless P = NP.

\begin{theo}
\label{th:noPTAS.high.d}
The Maximum TSP and the Maximum Scatter TSP are Max-SNP-hard for
points in $d$-dimensional space under $L_\infty$-norms when $d$
is part of the input.
\end{theo}

\proof
The source problem is
the Minimum TSP with all edge lengths in $\{1,2\}$, a special case that 
was proved Max-SNP-hard by Papadimitriou and Yannakakis \cite{py}.
We use the construction
of our previous proof, with edges of length 1 as ``edges,'' edges of
length 2 as ``non-edges,'' and each ``edge'' having its own coordinate
in $|E|$-dimensional space.  For this coordinate, one endpoint gets
value $+1$, the other gets $-1$, and all other points get value $0$.
Thus, adjacent vertices in $G$ get mapped to points at the ``long''
$L_\infty$-distance 2, while each pair of vertices at distance
2 in $G$ gets mapped to a pair of points at ``short'' $L_\infty$-distance 1.
Therefore, long tours in the constructed point set correspond to
short tours in the original graph. Now the Max-SNP-hardness is
immediate, as a Maximum TSP tour of length at least $(2-\eps)n$,
i.e., with at most $\eps n$ short edges,
corresponds to a Minimum TSP tour of length at most 
$(1+\eps)n$, i.e., with at most $\eps n$ edges of length 2.
\qed

\medskip
The question remains open for $L_p$, $1 \leq p < \infty$,
although we conjecture that these cases are Max-SNP-hard as well.

%Finally, we note that our results can be extended in several ways.
%For instance, one can get polynomial-time algorithms for asymmetric
%versions of the Maximum TSP in which distances are computed based on
%non-symmetric unit balls.
%Also, algorithmic approaches analogous to ours can be applied to
%geometric versions of other NP-hard maximization problems.
%For example, consider the {\em Weighted 3-Dimensional Matching Problem\/}
%that consists in partitioning a set of $3n$ elements into $n$ triples of 
%maximum total weight.
%The special case where the elements are points in
%$\Rd$ and where the weight of a triple equals the perimeter of the 
%corresponding triangle measured according to some fixed polyhedral norm 
%can be solved in polynomial time.

\section{Conclusion}
\label{sec:conc}
\nopagebreak
We have derived polynomial time algorithms for the Maximum TSP when the
cities are points in $\Rd$ for some fixed $d$ and when the distances
are measured according to some polyhedral norm or quasi-norm,
with running time
$O(n^{k-2}\log n)$ for norms based on $k$-facet polyhedra and
$O(n^{2k-2}\log n)$ for quasi-norms based on $k$-facet polyhedra. 
Our approach is based on a solution method for the Tunneling TSP;
we believe that the related Minimum TSP variant with city connections
via a fixed set of hubs is of independent interest.
We also gave an optimal
$O(n)$ algorithm for the special case of 4-facet polyhedra in the plane, such
as the rectilinear norm. 
We suspect it may be possible to improve
on our complexity for
$L_1$ distances in $\Rr$ by using some of our geometric ideas.
(Since the unit ball for $L_1$ distances in $\Rr$
is an octahedron, the running time for our general algorithm
is $O(n^6\log n)$.)
 
We have also shown that the Maximum TSP under Euclidean norm in $\Rd$
is NP-hard for any fixed $d\geq 3$. This shows that the complexity
of an optimization problem is not just a consequence of its
combinatorial structure or its geometry, but may be
ruled by the structure of the particular distance function that is
used. The result has similar implications for closely related problems.
 
The Euclidean case $d=2$ remains open; in the light of our results,
it seems more likely that this problem is NP-hard, even though
its counterpart with rectilinear distances turned out to be extremely simple.
However, it is much harder to
use strictly local arguments for geometric maximization problems,
so a proof of NP-hardness may have to use a more involved construction.
 
\begin{conj}
\label{r2}
The Maximum TSP for Euclidean distances in the plane is an NP-hard problem.
\end{conj}

\old{ 
A further difficulty with the Euclidean norm (one shared by both the 
Minimum and Maximum TSP) is that we still do not know whether the TSP is 
in NP under this norm.
Even if all city coordinates are rationals, we do not know how to compare
a tour length to a given rational target in less than
exponential time.
Such a comparison would appear to require us to evaluate a sum of $n$ 
square roots to some precision, and currently the best upper bound known 
on the number of bits of precision needed to insure a correct answer
remains exponential in $n$.
Thus even if we were to produce an algorithm for the Euclidean Maximum TSP
that ran in polynomial time when arithmetic operations (and comparisons)
take unit time, it might not run in polynomial time on a standard Turing 
machine.
}

\vspace{.1in}
\noindent{\bf Acknowledgment.} 
Thanks to Henk Meijer, Estie Arkin, Volker Kaibel, Joe Mitchell,
Bill Pulleyblank, Mauricio Resende, Peter Shor, and Peter Winkler
for helpful discussions, and to an anonymous referee for useful comments.


\begin{thebibliography}{4}

\bibitem{AOST}
Ahuja, R.K., Orlin, J.B.,  Stein, C.,  and Tarjan, R.E.,
``Improved algorithms for bipartite network flow," {\it SIAM J. Comp.}
{\bf 23}, (1994), 906--933.

\bibitem{ACMSY}
Arkin, E.M., Chiang, Y.-J, Mitchell, J.S.B., Skiena, S.S.,
and Yang, T.-C.,
\newblock ``On the Maximum Scatter TSP,''
\newblock {\em Proc.\ 8th ACM-SIAM Symp.\ Disc.\ Alg.\ (SODA 97)}, 
1997, 211--220.

\bibitem{arora1} 
Arora, S.,
``Polynomial-time approximation schemes for Euclidean TSP and other 
geometric problems,''
%{\em Proc. 37th IEEE Symp. on Foundations of Computer Science},
%IEEE Computer Society, Los Alamitos, CA, 1996, 2--12.
{\em J.\ ACM} {\bf 45}, (1998), 753--782.

%\bibitem{arora2} 
%Arora, S.,
%``Nearly linear time approximation schemes for Euclidean TSP and other 
%geometric problems,''
%{\em Proc. 38th IEEE Symp. on Foundations of Computer Science},
%IEEE Computer Society, Los Alamitos, CA, 1997, 554--563.

\bibitem{pcp} 
Arora, S., Lund, C., Motwani, R., Sudan, M., and Szegedy, M.,
``Proof Verification and Hardness of Approximation Problems,'' 
{\em J.\ ACM}, {\bf 45}, (1998), 501--555.

\bibitem{Baj88}
Bajaj, C.,
``The algebraic degree of geometric optimization problems,''
{\em Disc.\ Comp.\ Geom.}, {\bf 3} (1988), 177--191.
 
\bibitem{Barvinok} 
Barvinok, A.I.,
``Two algorithmic results for the traveling salesman problem,''
{\em Math.\ Op.\ Res.}, {\bf 21} (1996), 65--84.

\bibitem{BaJoWoWo98}
Barvinok, A.I., Johnson, D.S., Woeginger, G.J., and Woodroofe, R.,
\newblock ``The maximum traveling salesman problem under polyhedral norms,''
\newblock {\em Proc.\ 6th Int.\ Integer Prog.\ 
Comb.\ Opt.\ Conf.\ (IPCO VI)},
Springer LNCS 1412, 1998, 195--201.

\bibitem{Be83}
Ben-Or, M.,
``Lower bounds for algebraic computation trees,''
{\em Proc.\ 15th ACM Symp.\ Theory Comp.\ (STOC 83)}, 1983,
80--86.

\bibitem{Cayley}
Cayley, A.,
``A theorem on trees,''
{\em Quarterly Journal on Mathematics}, {\bf 23} (1889),
376--378.

\bibitem{CLRS}
Cormen, T.H., Leiserson, E.L., Rivest, R.L., and Stein, C.
{\em Introduction to Algorithms} (2nd ed.),
MIT Press, Cambridge, 2001, p.\ 195.

\bibitem{linear}
Blum, M., Floyd, R.W., Pratt, V.R., Rivest, R.L., and  Tarjan, R.E.,
``Time bounds for selection,''
{\em J.\ Computer Syst.\ Sc.}, {\bf 7} (1972), 448--461.
 
\bibitem{F99}
Fekete, S.P., 
``Simplicity and hardness of the maximum Traveling Salesman Problem
under geometric distances,''
{\em Proc.\ Tenth ACM-SIAM Symp.\ Disc.\ Alg.\ (SODA 99)}, 1999,
337--345.

\bibitem{FM}
Fekete, S.P., and Meijer, H.,
``On minimum stars and maximum matchings,''
{\em Disc.\ Comp.\ Geom.}, {\bf 23} (2000), 389--407.

\bibitem{FiSuTu97}
Fingerhut, J.A., Suri, S., and Turner, J.S.,
``Designing least-cost nonblocking broadband networks,''
{\em J.\ Alg.}, {\bf 24} (1997), 287--309.

\bibitem{GGJ}
Garey, M.R., Graham, R.L., and Johnson, D.S., 
``Some NP-complete geometric problems,'' 
{\em Proc.\ 8th ACM Symp.\ on Theory of Computing (STOC 76)} 1976, 10--22,

\bibitem{GJT}
Garey, M.R., Johnson, D.S., and Tarjan, R.E.,
``The planar Hamiltonian circuit problem is NP-complete,''
{\em SIAM J.\ Comput.} {\bf 5} (1976), 704--714.

\bibitem{GMF}
Gusfield, D., Martel,  C., and Fernandez-Baca, D.,
``Fast algorithms for bipartite network flow,''
{\em SIAM J.\ Comp.}, {\bf 16} (1987), 237-251.

\bibitem{7/8}
Hassin, R.\ and Rubinstein, S.
``A $\frac{7}{8}$-approximation algorithm for metric Max TSP,''
{\em Inf.\ Proc.\ Lett.}, {\bf 81} (2002), 247--251.

\bibitem{ItPaSw}
Itai, A., Papadimitriou, C., and Swarcfiter, J.L.,
``Hamilton paths in grid graphs,''
{\em SIAM J.\ Comp.} {\bf 11} (1982), 676--686.

\bibitem{TSP}  
Lawler, E.L., Lenstra, J.K., Rinnooy Kan, A.H.G., and Shmoys, D.B., 
{\em The Traveling Salesman Problem}, 
Wiley, Chichester, 1985.

\bibitem{Matsui}  
Matsui, T., ``Linear time algorithm for the Hitchcock transportation
problem with a fixed number of supply points," 
Optimization -Modeling and Algorithms-,
{\em Cooperative Research Report} {\bf 35} (1992), 
The Institute of Statistical Mathematics,
Minami-Azabu, Minato-ku, Tokyo, Japan, 128--138.

\bibitem{MeTa93}
Megiddo, N., and Tamir, A.,
``Linear time algorithms for some separable quadratic programming problems,''
{\em Oper.\ Res.\ Lett.} {\bf 13} (1993), 203--211.

\bibitem{mitch}
Mitchell, J.S.B.,
``Guillotine subdivisions approximate polygonal subdivisions: Part II --
A simple PTAS for geometric $k$-MST, TSP, and related problems,''
{\em SIAM J.\ Comp.}, {\bf 28} (1999), 1298--1309. 
 
\bibitem{joe.priv}
Mitchell, J.S.B.,
personal communication, 1998.

\bibitem{Pap}
Papadimitriou, C.H., 
``The Euclidean traveling salesman problem is NP-complete,'' 
Theoretical Comp.\ Sci.\ {\bf 4} (1977), 237--244.

\bibitem{py} 
Papadimitriou, C.H., and Yannakakis, M.,
``The traveling salesman problem with distances one and two,''
{\em Math.\ of Oper.\ Res.} {\bf 18} (1993), 1--11.

\bibitem{Serd1}
Serdyukov, A. I.,
``An asymptotically exact algorithm for the
traveling salesman problem for a maximum in Euclidean
space'' (Russian), 
{\em Upravlyaemye Sistemy} {\bf 27} (1987), 79--87.

\bibitem{Serd2}
Serdyukov, A. I.,
``Asymptotic properties of optimal solutions of
extremal permutation problems in finite-dimensional
normed spaces'' (Russian),
{\em Metody Diskret. Analiz.} {\bf 51} (1991), 105--111.

\bibitem{Serd3}
Serdyukov, A. I.,
``The Traveling Salesman Problem for a maximum in
finite-dimensional real spaces'' (Russian),
{\em Diskret. Anal. Issled. Oper. 2}, {\bf 1} (1995), 50--56.

%\bibitem{Suri}
%Suri, S.,
%Problem \# 5,
%Problem session of the
%{\em 14th ACM Symposium on Computational Geometry}, 1998.

\bibitem{TaMi98}
Tamir, A., and Mitchell, J.S.B.,
``A maximum $b$-matching problem arising from median location models with applications to the roommates problem,''
{\em Math.\ Prog.}, {\bf 80} (1998), 171--194.


\bibitem{TN} 
Tokuyama, T., and Nakano, J., ``Efficient algorithms for the Hitchcock
transportation problem,'' {\it SIAM J. Comp.}
{\bf 24}, (1995), 563-578.

\bibitem{Trev} 
Trevisan, L.,
``When Hamming meets Euclid: The approximability of geometric TSP and MST,''
{\em Proc.\ 29th ACM Symp.\ Theory Comp. (STOC)},
ACM, New York, 1997, 21--29.

\bibitem{Zemel} 
Zemel, E.,
``An $O(n)$ algorithm for the linear multiple choice knapsack problem and
related problems,''
{\em Inf.\ Proc.\ Lett.} {\bf 18} (1984), 123--128.
\end{thebibliography}
\end{document}